# The Dark Energy Survey data management system

Joseph J. Mohr[a,b], Darren Adams[b], Wayne Barkhouse[c], Cristina Beldica[b], Emmanuel Bertin[d], Y. Dora Cai[b], Luiz A. Nicolaci da Costa[e], J. Anthony Darnell[a], Gregory E. Daues[b], Michael Jarvis[f], Michelle Gower[b], Huan Lin[g], Leandro Martelli[e], Eric Neilsen[g], Chow-Choong Ngeow[a], Ricardo L. C. Ogando[e], Alex Parga[b], Erin Sheldon[h], Douglas Tucker[g], Nikolay Kuropatkin[g], and Chris Stoughton[g] for the Dark Energy Survey Collaboration

[a]University of Illinois Department of Astronomy, 1002 West Green St, Urbana, IL, USA 61801;
[b]National Center for Supercomputing Applications, 1205 West Clark St., Urbana, IL, USA 61801;
[c]University of North Dakota Department of Physics, 101 Cornell St., Grand Forks, ND, USA 58202;
[d]Institut d'Astrophysique, 98bis, bd Arago, Paris, France 75014;
[e]Observatorio Nacional, R. Gal. Jose Cristino, 77, Rio de Janeiro, RJ, Brazil;
[f]University of Pennsylvania Department of Physics, 203 South 33rd St., Philadelphia, PA, USA 19104;
[g]Fermi National Accelerator Laboratory, P. O. Box 500, Batavia, IL, USA 60510;
[h]New York University Department of Physics, 4 Washington Place, New York, NY, USA 10003;

**ABSTRACT**

The Dark Energy Survey (DES) collaboration will study cosmic acceleration with a 5000 deg$^2$ *griZY* survey in the southern sky over 525 nights from 2011-2016. The DES data management (DESDM) system will be used to process and archive these data and the resulting science ready data products. The DESDM system consists of an integrated archive, a processing framework, an ensemble of astronomy codes and a data access framework. We are developing the DESDM system for operation in the high performance computing (HPC) environments at the National Center for Supercomputing Applications (NCSA) and Fermilab. Operating the DESDM system in an HPC environment offers both speed and flexibility. We will employ it for our regular nightly processing needs, and for more compute-intensive tasks such as large scale image coaddition campaigns, extraction of weak lensing shear from the full survey dataset, and massive seasonal reprocessing of the DES data. Data products will be available to the Collaboration and later to the public through a virtual-observatory compatible web portal. Our approach leverages investments in publicly available HPC systems, greatly reducing hardware and maintenance costs to the project, which must deploy and maintain only the storage, database platforms and orchestration and web portal nodes that are specific to DESDM. In Fall 2007, we tested the current DESDM system on both simulated and real survey data. We used Teragrid to process 10 simulated DES nights (3TB of raw data), ingesting and calibrating approximately 250 million objects into the DES Archive database. We also used DESDM to process and calibrate over 50 nights of survey data acquired with the Mosaic2 camera. Comparison to truth tables in the case of the simulated data and internal crosschecks in the case of the real data indicate that astrometric and photometric data quality is excellent.

## 1. Introduction

Several independent lines of observational evidence indicate that the expansion of the universe has entered an accelerating phase[1,2,3,4]. A key objective in cosmology and high-energy physics today is to understand the underlying cause of this comic acceleration. It can be explained by a ubiquitous dark energy component that dominates the current energy density of the universe; the observed acceleration can also be explained by a flaw in our understanding of the behavior of gravity on the largest scales. Several promising astrophysical methods for studying the cosmic acceleration have been developed, and each draws upon distance measurements extending over cosmological scales and/or studies of the growth rate of cosmic structures. Measurements of either distances or the growth rate of structure constrain the expansion history of the universe *H(z)* (Hubble parameter as a function of redshift). The DES Collaboration is pursuing precise measurements of the cosmic expansion history employing four independent techniques: (1) galaxy cluster surveys, (2) cosmic shear, (3) clustering of galaxies, and (4) SNe distance estimates. Each of these methods is subject to different systematic errors, so by building all four measurements into a single experiment we will obtain important cross checks on the constraints. Because each method measures the expansion history in a different way, the measurement constrains a different combination of properties of the universe (i.e. geometry, dark energy density, dark energy equation of state parameter, etc.); therefore, combined constraints from the four different techniques enable a far more precise measurement of the properties of dark energy, including test of the underlying dark energy hypothesis. To achieve these science goals, the DES collaboration will undertake a deep, 5 band survey of 5000 deg$^2$ of sky and repeatedly image a smaller region to identify and measure light curves for supernovae. This survey requires a new wide field camera (DECam) for the existing Blanco 4m telescope together with the software required to transform the survey data into the products needed for the science analyses.

The DES data management (DESDM) project is developing the system required to process the raw DECam images into science ready data products and make those data available to the Collaboration and the community. The DESDM system will be used to process and archive approximately 200TB of raw imaging data into a few PB of science ready data products. Under the leadership of NCSA/UIUC, the DES Collaboration is developing, and will deploy and operate the DESDM system; this system consists of (1) the astronomy codes required to process the data, (2) pipelines with built in quality assurance testing, (3) a distributed archive to support automated data processing and calibration within a high performance computing environment, (4) a catalog database to support calibration, provenance and science analyses, (5) web portals for control, monitoring, user data access and scientific analyses, and (6) the hardware platforms required to operate the system. NCSA/UIUC will be the primary DES Archive and processing center, but the DESDM system has been designed to enable automated distribution of raw and reduced data products throughout the DES Collaboration.

The design of the DESDM system is driven by the scientific goals of the DES project as well as by the need to automatically and reliably process, calibrate and archive the survey dataset. Our objective is to meet or exceed the data quality requirements established in the DES Science Requirements Document and flowed down to the DESDM Requirements and Specifications document. The DESDM development plan follows a spiral or iterative model with periods of design refinement and development followed by periods of intense testing and validation, which we call Data Challenges. The advantages of this approach are that it ensures the resulting system meets the needs of the users, and that it enables us to validate architecture decisions and software package choices throughout the development phase. We have successfully conducted our first three Data Challenges (Oct '05, Oct '06, Oct '07) where we tested prototype versions of our DESDM system by processing simulated DECam data and Mosaic2 data from the Blanco Cosmology Survey. In our last data challenge we tested our DESDM system using a quite realistic dataset composed of 500 deg$^2$ of mock DECam data four layers deep and in four bands. This is equivalent to 8,000 deg$^2$ of imaging data, fully 20% of the entire SDSS imaging dataset. The reductions of the Mosaic2 data using DESDM are feeding scientific analyses within the Blanco Cosmology Survey and the South Pole Telescope collaborations.

In section 2 we describe the DESDM system, and in Section 3 we summarize our latest round of testing and validation. Section 4 contains a description of a quality assurance framework that will operate in a distributed computing environment, and Section 5 describes our work to integrate the DESDM system with the science analysis codes. Finally, in Section 6 we conclude with a summary of our management structure, funding sources and development plan.

## 2. DESDM Processing System

To achieve the DES science goals, the DESDM system and team must (1) process a huge, multiband optical survey into science-ready images, catalogs and metadata, (2) archive and distribute these data products to the collaboration and ultimately to the public, and (3) enable the scientific analyses of these data by the DES collaboration and the public. Figure 1 provides a graphical representation of the system.

Data move by network from CTIO to NCSA/UIUC using the existing NOAO Data Transport System. Receiving a long night of data (360 exposures) in 18hrs requires an average bandwidth of 36Mbps. The raw data arrive at NCSA/UIUC where they are ingested into the DES Archive, which includes an Oracle relational database. After data ingestion, an orchestration node, which is ultimately controlled by an operator through a portal, prepares and submits jobs to the pre-assigned HPC queues at NCSA and initiates any required data transfer jobs to position the raw data on the compute nodes for processing and retrieve the data products for safe storage. For nightly processing we will use scheduled time on a fixed number of nodes (i.e. 100), and so these queue submissions will flow immediately onto the compute nodes. Results from the processing are ingested into the DES Archive; this ingestion includes inserting cataloged object data into database tables as well as the file metadata describing the archive location and relevant properties of each image and catalog file. Quality assurance (QA) testing is integrated into the processing, and results of this testing will be available to the operators through a QA portal. The QA testing will be used to identify and respond to problematic datasets and computing failures. QA also produces metadata that are inserted into the DES Archive database to enable filtering of the data using quality flags.

We employ data parallel processing to enhance the throughput of our system. In our last data challenge, it took approximately 6 wall clock hours to complete the processing of a night (~300 science images), although the total CPU time is approximately 700 hours. Code enhancements and additional processing steps are being developed that will improve the data quality and that will prepare additional data products for input to the key science analyses. We will respond to these increases in computing needs by introducing higher levels of data parallel subdivision than we currently employ.

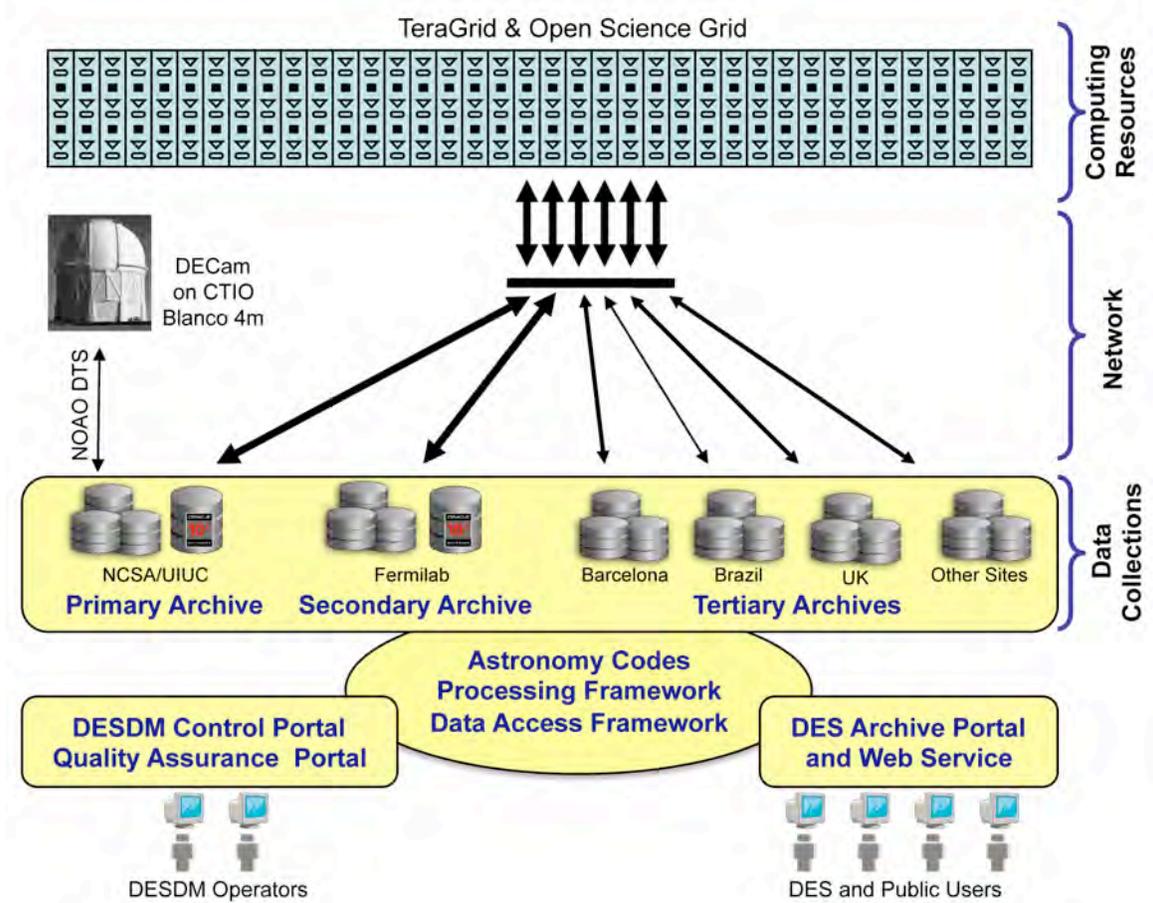

Figure 1: The DESDM System transforms DECam data into science ready data products using community high performance computing resources. Image and catalog data are tracked within a distributed archive, and data movement among archive nodes employs high efficiency grid tools. Operators control the system, submit processing requests and monitor data quality using portals. Collaboration scientists and the public retrieve data using the archive portal or web service. DESDM specific hardware includes the database and portal platforms together with the disk and tape data storage.

In addition to the nightly processing we expect to carry out bulk reprocessing between observing seasons. Full reprocessing of a single season of data will require approximately 8 CPU-years and produce approximately 300TB of data products, including approximately 4 billion objects for ingestion and calibration. With the DESDM system we can accomplish this reprocessing on a timescale of a week by using about 500 HPC nodes. Given the HPC resources at NCSA and Fermilab (as well as others within the collaboration), accessing these computing resources will not pose a significant challenge. Operating within the NCSA or Fermilab computing environment requires that our DESDM system be grid-enabled. Our approach offers significant cost savings to the project by leveraging public investment in

HPC resources, and it offers the flexibility to reach out for massive amounts of computing resources when they are needed. In addition, it brings a high level of robustness during nightly processing, because we can move to backup computational resources whenever problems arise with our preferred system.

In addition to these components, the DESDM system includes user interfaces for discovering, analyzing and retrieving data. Users from the DES Collaboration and the public will access data using either the DES web portal or a web service. We have taken steps to make our user sign-on compatible with the virtual observatory (VO) model, and we plan to continue to make our system VO compatible as standards are developed.

In the sections below we provide additional information about the astronomy codes, the processing framework and the archive and data movement tools.

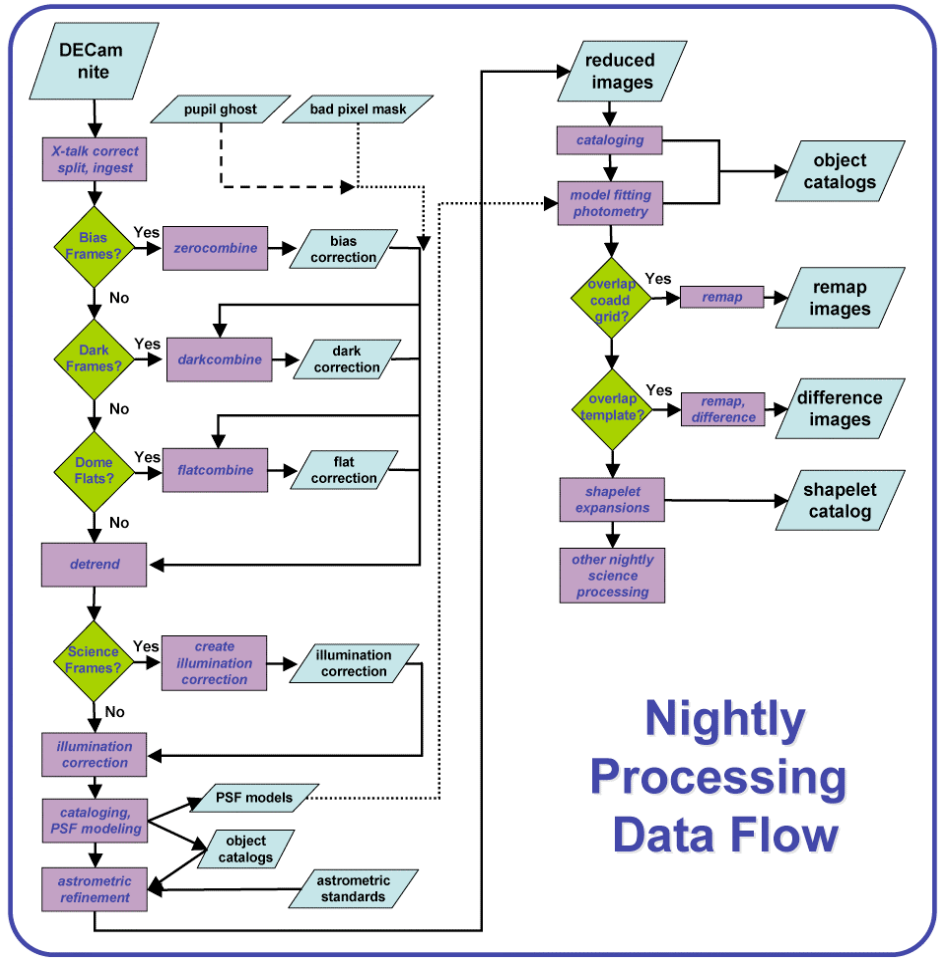

Figure 2: Nightly Processing includes image detrending, astrometric calibration, relative photometric calibration, cataloging with PSF corrected model fitting photometry, preparation of remapped images for later coaddition and production of difference images for cataloging of time variable sources All image and catalog data products are ingested into the DES Archive and database. Photometric calibration is carried out using database procedures.

### 2.1. Astronomy Codes

The astronomy codes are the core building blocks of our processing system. Figure 2 contains an overview of the steps in our nightly processing. Incoming FITS images from DECam are detrended using codes that we have developed. This detrending includes the standard crosstalk correction, pupil ghost correction, overscan correction, trimming, bias subtraction, flat fielding and illumination correction. We have also written the codes that create the calibration images required for each of these corrections.

After the image detrending, the images are astrometrically calibrated, remapped for later coaddition and cataloged. For photometric data, a photometric calibration is applied to the single epoch object photometry after the catalogs are ingested into the DES Archive database. A guiding philosophy for the Nightly Processing is to carry out as many preparatory calculations as possible to lighten the computing load on the higher level processing and analyses that rely on data collections spanning many observing nights. For example, we remap images, model the PSF and its variation over each image and execute shapelet expansions on objects during nightly processing; these data products are stored in the DES Archive for later use by the coaddition, weak lensing and difference imaging pipelines.

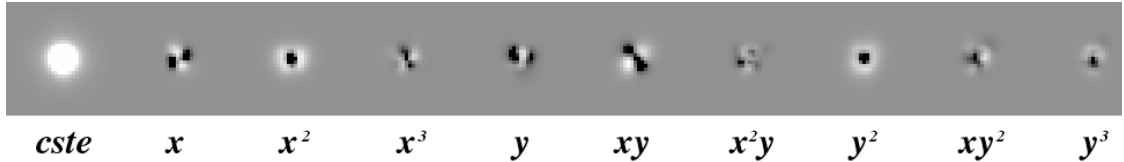

Figure 3: PSF components derived by PSFEx, a Terapix code that models the PSF variation over each image using a polynomial expansion. DESDM employs PSFEx to create a position dependent model in all detrended single epoch images to support PSF-corrected model fitting photometry, optimal star-galaxy classification and extraction of the weak lensing shear field.

We have adopted the Terapix software SExtractor, SCAMP, SWarp and PSFEx for cataloging, astrometric refinement, remapping for coaddition and modeling of PSF variations over each image. These codes are all developed by Dr. Emmanuel Bertin, who is leading the effort to extend these codes, where needed, to meet the DESDM requirements. A focus of his development at present is the implementation of model fitting photometry for galaxies. This approach relies on a local model of the PSF, and it allows for PSF fitting photometry of stars and two dimensional model based estimates of galaxy photometry and morphological characteristics. Figure 3 contains components of the PSF model that vary as polynomials of CCD (x,y) position.

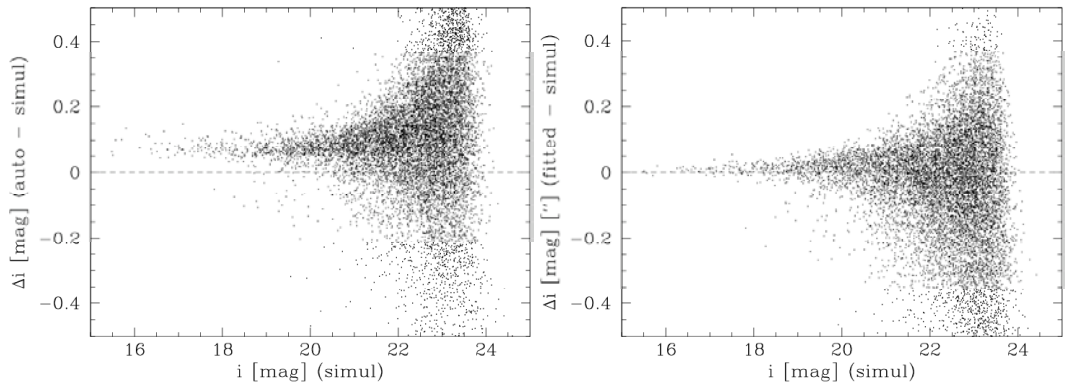

Figure 4: Comparison of *mag_auto* photometry (left) to model fitting photometry using SExtractor and simulated galaxy images. Note the significant reduction in photometric bias that is delivered by the model fitting photometry when compared to the Kron-like estimator mag_auto.

Figure 4 shows a test of SExtractor with the model fitting extension. In comparison to the traditional Kron-like magnitudes such as mag_auto, the model fitting photometry captures a much larger fraction of the total galaxy light. The resulting galaxy models in different bands allow one to estimate galaxy color profiles, enabling the possibility of multiple, independent photo-z estimates for well resolved galaxies. Moreover, we expect this approach to offer significant advantages for deblending in crowded fields (i.e. galaxy cluster cores). Finally, the local model of the PSF in every image allows for direct likelihood estimation of the probability that each object is resolved, providing an optimal approach for star—galaxy classification. The model fitting photometry is much more computationally taxing than traditional cataloging, but within our flexible HPC enabled system we can easily reach out for the additional resources required.

Another focus of our effort is the production of code to model the survey depth and completeness as a function of position. Completeness and depth are sensitive to both sky noise and seeing; we track the sky noise for each pixel, and we model the PSF variation over each exposure. DES coadd images of a particular piece of the sky will generally come from different nights under different observing conditions. If these images are directly coadded to build deeper images of the sky, the PSF will change discontinuously with position within the coadd images, complicating the process of modeling the survey depth and of extracting robust, model-fitting photometry of the objects. By applying a homogenization process that brings all input images to a common PSF before the coaddition, one produces coadd images where the catalog depth and completeness information is much easier to extract. We have developed tools to carry out PSF homogenization to the median seeing on the input single epoch images during the process of coaddition to explore whether the completeness and depth are then simply related to the coadd image pixel variances. One concern is that the homogenization process correlated the noise. Tests are currently underway, and our goal is employ a homogenization strategy in production during the next data challenge.

An additional challenge is in the production of accurate catalogs that rely on the coaddition of images from detectors with varying color terms (both within and among detectors). DECam detectors, optics and filters are being designed to minimize these variations, but ultimately these effects may provide a noise floor in the coadd photometry that can only be overcome with more sophisticated processing. For bright objects one could simply combine the independent, color corrected photometry extracted from all the single epoch images of that object. For fainter objects a more sophisticated algorithm that tackles the detector sensitivity variations directly during the coadd is likely required. We are currently exploring algorithms to overcome these challenges.

## 2.2. Processing Framework

Data processing is organized through an orchestration layer that is deployed on a minimum of two orchestration nodes to enhance the reliability of the system. The orchestration layer enables high levels of automation and robustness by helping the operator (1) determine what processing needs to be done, (2) assemble the necessary data, metadata and pipelines, and (3) submit processing jobs to compute platforms (whether local or remote). The orchestration layer also includes tools to monitors running jobs, automatically checking their status, capturing logs from the pipelines, and managing alerts generated by data QA tests. When the processing is complete, the orchestration layer triggers the ingestion and safe storage of the resulting data products (i.e. copying them to the NCSA mass storage system as well as to the primary and secondary DES Archive filesystems).

The pipelines themselves are astronomy processing modules organized into workflows using middleware that has been developed at NCSA. This middleware is the Open Grid Runtime Engine scripting language Elf/OgreScript. One important feature of Elf is an event channel, which provides a way to send and receive status, data quality and control messages using a central event repository. This layer plays an important role in process monitoring and QA, because data rich events can be published by the pipelines from anywhere on the grid and then queried by the orchestration layer. Our control and QA portals will also query this notification service and build summaries of the processing progress for inspection by the operator. This will be key for ensuring data quality and simplifying operations during the survey processing.

Our processing framework is designed to enable us to leverage industry leading grid computing software like the Globus Toolkit, Condor-G and developing tools like Elf/OgreScript. Going this route rather than writing our own special purpose grid computing tools allows us to (1) concentrate our efforts on the more specialized astronomy components of the DESDM system, (2) benefit from the many strengths of these industry leading components and the future improvements to these tools, and (3) harness the computing power of publicly available NSF and DOE supported computing facilities (in contrast to being "hard wired" to work on a specific cluster available to only one project).

The DESDM system operator will control the system through a portal. We have developed and tested a prototype portal, which supports the basic detrending processing of the DECam images. Additional design and development is ongoing with the goal of including the following functionality: (1) select dataset(s) for processing, (2) select modules within our library that will be combined on the fly into a pipeline, (3) adjust parameters within the selected modules, (4) select HPC compute platforms, (5) prestage data and submit the pipeline(s) for processing on those platforms, and (6) select destinations for automated safe storage or retrieval of the data. The DESDM system we are testing today supports all these actions, and so control portal development is focused on constructing a simple graphical interface that allows the operator easy access to this existing functionality.

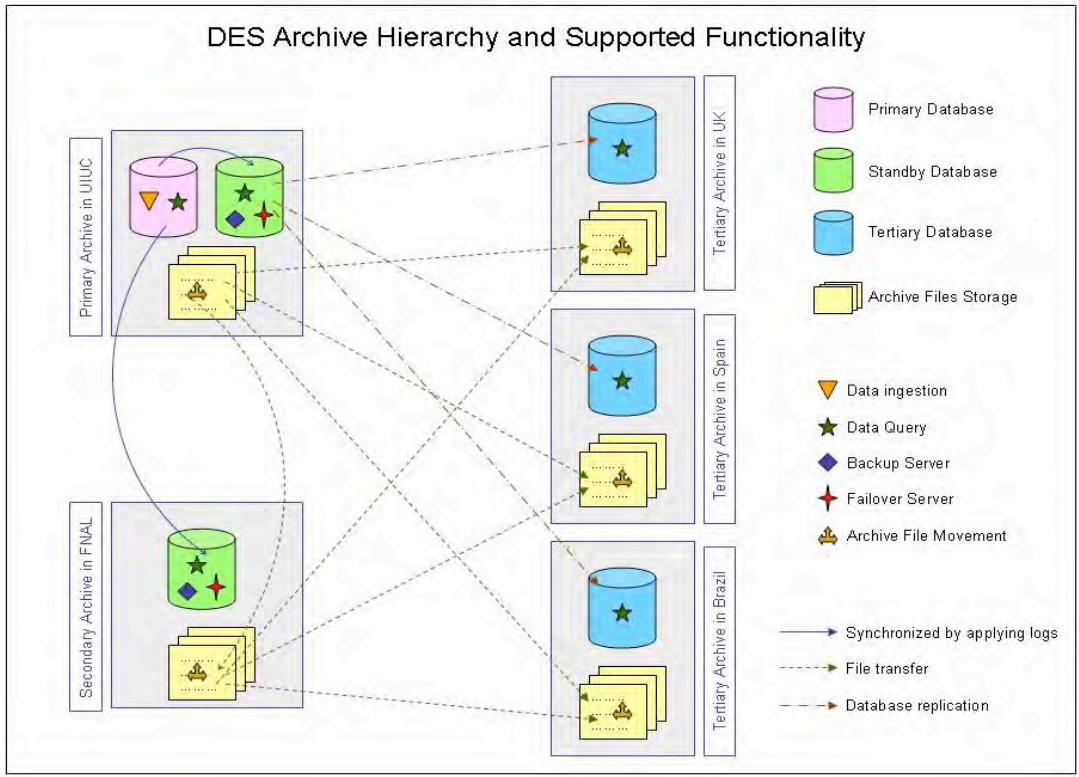

Figure 5: The DES Archive is composed of an Oracle database and filesystem storage that will exceed 1PB over the life of the survey. We have designed a doubly redundant database that enables continuous operation during routine software upgrades as well as during platform failures or site failures at UIUC, the Primary Archive site, when operations will shift to Fermilab, the Secondary Archive site. Tertiary archive sites at many DES institutions will provide local access to the latest science-ready data products, and databases at those locations will be employed by a query-broker service to support science queries by the Collaboration and the public.

**2.3. DES Archive and Data Access Framework**

The DES Archive is composed of archive nodes (data collections within the filesystems on various hosts), a relational database, and a data access framework (DAF); metadata in the relational database enable tracking of data across multiple nodes, and the DAF facilitates the transfer of data products among these nodes. The archive is designed to support the processing framework, which employs remote queries to discover the input data for processing and remote ingestion to archive the resulting products. The DES Archive is designed to track data on multiple archive nodes, which provides critical flexibility within our HPC processing environment and enables simple, automated data delivery within the DES Collaboration. Figure 5 shows the relationships among the Primary, Secondary and Tertiary archive sites. UIUC/NCSA will be the primary archive site that supports the bulk of the data processing and calibration. Fermilab will be the secondary archive site, hosting a complete dataset to enable processing failover during periods of software or hardware upgrades or unexpected disruptions. We expect many DES institutions will build Tertiary archive sites to enjoy local access to the data. Data will be available to users through an Archive Portal and an Archive Web Service. We intend to make the DES Archive data access interfaces VO-compatible to support DES data discovery and manipulation through other VO portals.

At the primary archive and processing center, we will deploy approximately 1PB of spinning disk over the life of the survey, with about 10% of that dedicated to support the DES Archive database. The NCSA mass storage system will play the role of full data safe store. User or pipeline access to data on disk or on tape will differ only in the latency between data request and the beginning of the data download. Below are more detailed descriptions of the Archive components.

### 2.3.1. DES Archive Nodes

The archive nodes are the repositories for all DES data products as well as logs, quality assurance reports, scripts, configuration files, pipelines and executables associated with the processing jobs. All archive nodes employ the same logical directory structure for data, and this makes it straightforward to uniquely identify any resource in the archive with a small number of logical identifiers or metadata tags. At present we store up to 14 tags for every file in our Archive, including the file creation date and size. We are exploring the option of storing file checksums for all archived files rather than just FITS images and catalogs. There are approximately $10^5$ files for each night of processed data with the current version of the DESDM system, implying that our system must support upwards of $10^9$ files at survey end when reprocessing and distribution to multiple archive nodes is included.

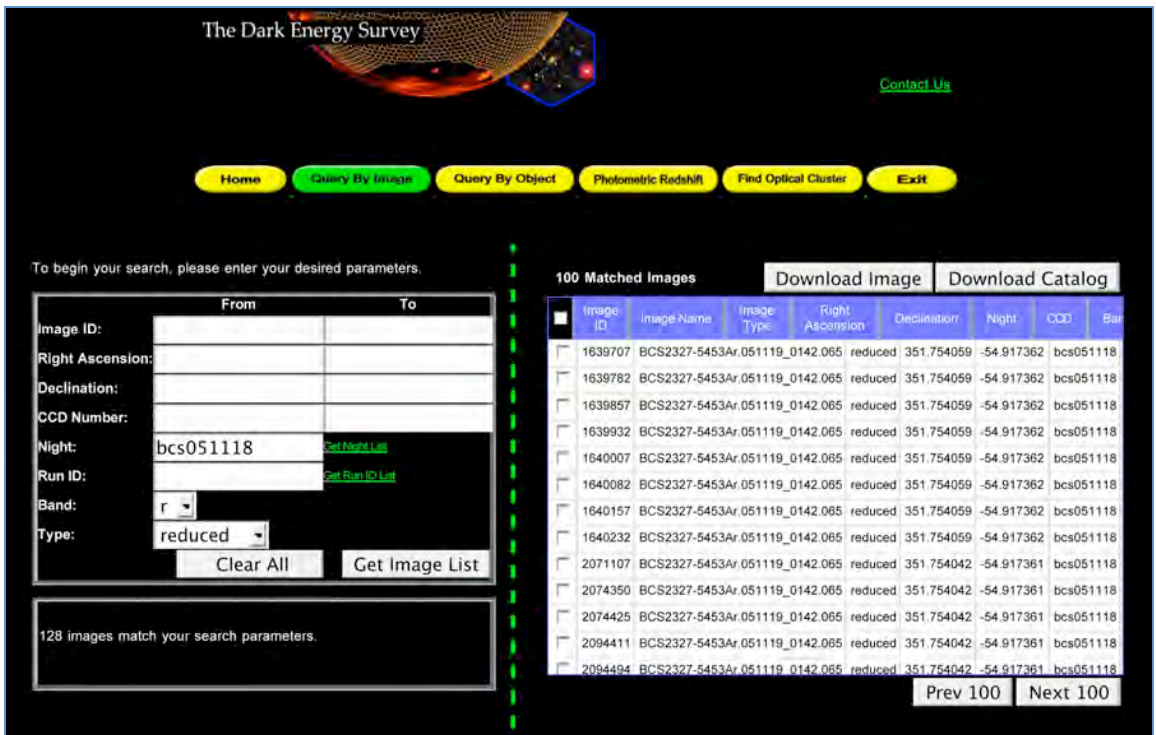

Figure 6: Screenshot of current prototype of the DES Archive web portal.

### 2.3.2. DES Archive Database

The DES Archive database is designed to support the DESM processing framework, to support science access to the Archive, and to provide an important analysis tool for the Collaboration and the public. We have chosen the Oracle relational database to meet these needs. Currently the DES Archive database includes 28 tables with over 70 indices to enable efficient linking of data across tables and 50 stored procedures, including several high performance matching algorithms that simplify object association using sky coordinates. Several of these tables contain object catalog data, including our primary OBJECTS table, which we expect to grow to ~100 billion rows by survey end. Other tables contain image science metadata, file location metadata, metadata describing the survey and the way we analyze the survey data, astrometric and photometric calibration tables, a software versioning table and many associated catalogs required to support the science analysis. Table and indices are designed to support data provenance through the linking of each object back to its image of origin, date of production, software stack used in production and ultimately back to the raw exposure from which that image was derived. The operational size of our database by survey end will be approximately 100TB.

Our development plan includes extensive testing of the database as both an integral part of the processing system and a powerful tool for scientific analysis and discovery. Through this testing we have now adopted a table partitioning and partition merging scheme for our large objects table that has multiple indices. This approach has dramatically improved the scaling of database ingestion while maintaining index-based query performance that is almost independent of table size. During testing we are using an 8 core server with multiple, internal hardware RAID controllers and 32GB of RAM. We are planning an Oracle Real Application Cluster (RAC) system for operations, which should boost performance significantly by enabling access to the same database from multiple, independent (multi-core) compute servers that share a fiber channel network of RAID arrays, each consisting of a hardware RAID controller and a dozen or more SATA drives. This system provides the flexibility to grow both the database volume by adding additional RAID arrays and the database performance by adding additional compute servers.

Because the DES Archive database plays a vital role in the DESDM system, we have developed a standby/failover architecture that ensures good and continuous performance. At the primary archive site we will have two database platforms running in primary/standby configuration. This allows the database backup load to be shifted to the standby platform. The standby database also will serve as the failover database, should there be glitches on the primary database. In addition, the secondary database to be deployed at Fermilab will function as a standby/failover system that will support operations during UIUC site failures as well as during software or hardware upgrades.

### 2.3.3. Data Access Framework

Our data access framework integrates high performance grid data movement tools (like uberftp, globus-url-copy and Trebuchet) with the data tracking capabilities of the Archive to create simple to use tools for moving data among Archive nodes. As an example, *arcp* takes input from the user (or orchestration layer) that describes the data to be moved using the logical identifiers presented above along with source and destination Archive nodes. If we are moving source data we would identify it using the *nite* identifier for the night the data were acquired, and if we are moving processed data products we would identify it using the *run* identifier. *arcp* proceeds by querying the DES Archive database to map the archive node names to (1) the grid-ftp servers associated with the nodes, (2) the preferred grid-ftp transfer tool, and (3) the optimal parameter set (i.e. buffer size, number of streams) for that technology. Then *arcp* queries the database to create a list of files for transfer and initiates the grid-ftp transfer. After the transfer has completed successfully, *arcp* updates the file metadata in the DES Archive database to reflect that these data are now present on the destination node. Reliable file transfer is ensured through comparison of file sizes on source and destination nodes, and copy jobs that fail in the middle will restart efficiently at the appropriate point in the transfer. In addition to *arcp* there are several other related tools to support data removal and verification. These tools have been critical for the large scale data movement required in our second and third Data Challenges.

### 2.3.4. Archive Web Portal and Web Service

Users will retrieve image and catalog data from the DES Archive using the web portal and a web service. Figure 6 contains a screenshot of the working prototype that takes user data selection criteria, uses those to search the DES Archive database for all matching products, allows users to select from those products, and then delivers those data directly to the user with an ftp download. The web service that we are designing will employ tools that are also used for the web portal, but user data selection parameters will be encoded in the URL rather than entered into the portal web page. The web service will be especially useful in the automated generation of control or QA portal views that include links to data. For example, links to log files containing more complete information generated during a processing run will be a single click away for DESDM operators through an embedded call to the Archive web service.

User access to the Archive through the portal or the web service will be controlled using the identity on an X.509 certificate generated by the VO certifying authority. Currently, users coming to our portal are redirected to the NCSA VO certifying authority for login, where an appropriate certificate is generated. Users arriving with a valid certificate generated at any VO portal will be mapped onto access permissions in our DES Archive. Users must register once with the DES Archive to establish the correct data access rights to our data. Unregistered users are provided access only to publicly available data. Currently we are using this portal to support Collaboration examination of the DES data challenge products and to host the Blanco Cosmology Survey dataset.

## 3. Testing and Validation

In our development plan we undertake large scale tests or data challenges of the most current version of the DESM System each fall (October through January). During these tests we process large quantities of simulated DECam data produced at Fermilab. Ultimately, the DESDM system will be subject to acceptance tests arranged by the DES Science Committee, and then by the CTIO Director as part of the commissioning and integration of the full DECam and DESDM systems. Here we describe the testing and validation during the development period.

Our data challenges start in October of each year and end with an examination of the performance of our system with respect to the DESDM requirements. Each evaluation of the current system is used to refine the development plan for the following year. This is a model that has been widely applied within the software development community. The spiral development strategy offers significant risk mitigation for our project, because we have a working DESDM system in place very early in the process, and then we focus on improving components and adding functionality.

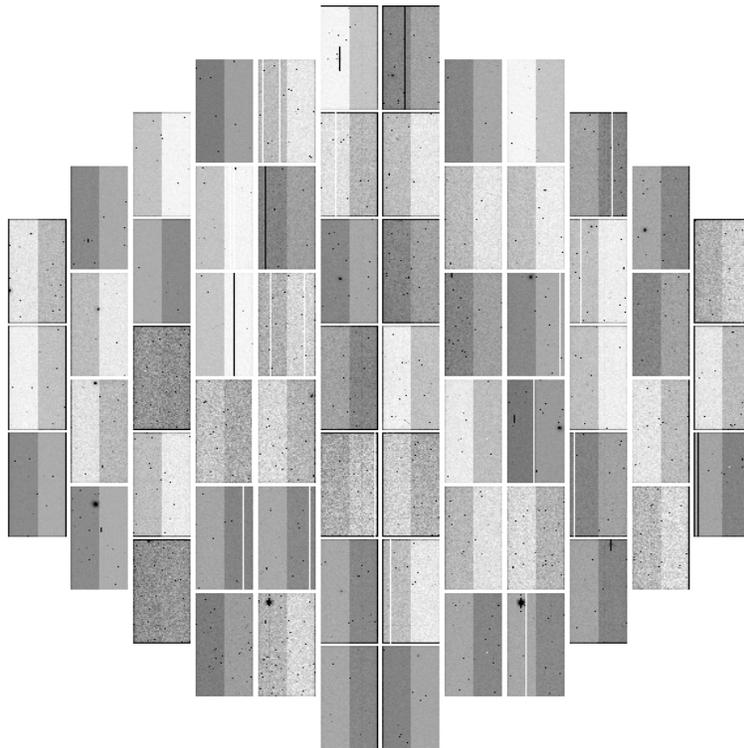

Figure 7: An example of a simulated DECam exposure with 62 science CCDs (2048x4096). Each CCD has two amplifiers, typically with different bias levels.

### 3.1. Simulated and Real Data

Each of the data challenges involves reducing simulated DECam data that are made increasingly realistic over time. In addition, there is a ramping up of the *volume* of imaging data to the point where we carry out a stress test of our system using a full simulated season of observing. These simulated data are produced at Fermilab by the DES Simulation team under the direction of Dr. Huan Lin. The mock DECam images include a wide range of realistic effects: (1) seeing variations that reflect the measured distribution on the Blanco + Mosaic2, (2) field distortions and spot sizes calculated using the design of the DECam corrector, (3) bad pixels, bad columns and edge effects consistent with the specifications of the DECam detectors, (4) cosmic rays, (5) pupil ghost at the 5% level, consistent with expectations given the optical system, (6) read noise and gain variations consistent with our expectation, (7) sky brightness variations due to lunar phase, (8) pointing errors and differential refraction across the field, and (9) modeling of non-photometric nights. Other effects are being added. The actual input catalog data include stars from the USNO-B catalog (important for astrometric calibration) together with galaxies sampled from large scale structure formation simulations. These galaxies are appropriately clustered and sheared in a manner consistent with the simulated matter distribution along the line of sight. In addition to the simulated images, the Fermilab Sim team provides truth tables (tables containing the input positions, brightnesses and morphological parameters for each object), which we ingest into the DES Archive database and use in our system tests.

We also test our system using the Blanco Cosmology Survey[5] (100 deg$^2$ *griz* survey) Mosaic2 data. Processing real data has been extremely valuable to our effort, and as the BCS science progresses it will serve as the first scientific product for the DESDM System.

### 3.2. Data Challenges

The data challenges involve using our DESDM system to reduce simulated and real data. This process includes (a) data delivery, (b) image data reduction to the catalog level, (b) photometric zeropoint estimation, (c) co-adding and cataloguing of the co-added images, and (d) production of final catalogs. These catalogs are then compared to the truth tables, enabling us to obtain information about accuracy of our photometry, astrometry, photometric redshifts and galaxy morphologies. In addition, these catalogs are then made available through the DES Archive web portal to the DES science working groups for scientific analysis, which will enable end to end tests of our ability to find galaxy clusters and estimate redshifts, recover the clustering of the galaxies, recover the weak lensing shear distribution and measure light curves of SNe Ia. We carried out a preliminary data challenge (called DC0) in spring 2005, when we altered two existing pipelines to process mock DECam data into object catalogs. Thereafter, we have maintained a fall testing schedule for DC 1-3, beginning in October '05, '06 and '07.

### 3.2.1. Processing of SDSS-scale Simulated Datasets

In DC3 we demonstrated that then current version of the DESDM System was already up to the challenge of processing, calibrating and archiving ten nights of DES data, equivalent to 8000 deg$^2$ or 20% of the full SDSS imaging dataset. The processing was carried out on the TeraGrid high performance computing platforms (primarily the Mercury cluster at NCSA and the IA64 cluster at SDSC), and the image and catalog data products were transferred to a dedicated 30TB repository on GPFS-WAN at SDSC. We also carried out our first tests of the DESDM Nightly Processing pipelines on the Open Science Grid (OSG), processing a night of DC3 data on a portion of the FermiGrid. All data movement used high bandwidth grid tools integrated within the DESDM data access framework. In our Nightly Processing we ingested and photometrically calibrated over 393 million objects and over 1.3 million images and catalogs.

Several higher level processing pipelines were tested as well, including the coadd pipeline, which we used to build and cataloging approximately 50 tiles, only a small fraction of the 1500 tiles we could have constructed with the full DC3 simulated dataset. Initial tests of components of the weak lensing pipeline were carried out. Shapelet expansions of objects and PSF standards were extracted during the nightly processing, and these expansions were ingested into the DES Archive database to support the shear extraction in later stages of the pipeline.

Catalog data quality was generally excellent. Comparisons of our cataloged objects and truth tables indicate we have achieved an astrometric systematic error floor of 20 mas when processing each exposure independently. Terapix experience with other datasets suggests that if we calibrate large ensembles of overlapping exposures together we may be able to reduce this floor by as much as a factor of four, and this may have benefits for portions of the DES weak lensing analysis. Stellar and galaxy photometry performance with SExtractor is quite good on the single epoch images, but systematic effects leading us to miss small amounts of light in faint objects cropped up in the deeper, coadd images. We have developed a strategy for correcting this problem, and small scale tests lead us to expect to be able to be able to demonstrate that we have achieved our 2% photometry requirements for non-crowded fields in the DC4 tests we will carry out in fall '08.

We learned a great deal from DC3. These tests have led to a significant refinement of the Processing Framework that will simplify the operation of the DESDM System. Moreover, we carried out scaling tests on the ingestion and query performance of the DES Archive database that have led us to partition our largest tables and employ a partition merging scheme during ingestion. Finally, our DC3 tests of galaxy and stellar photometry led us to plan additional refinements of the DECam simulations so that they better capture the faint halos surrounding galaxies (and to some extent, stars) that are seen in the real BCS images.

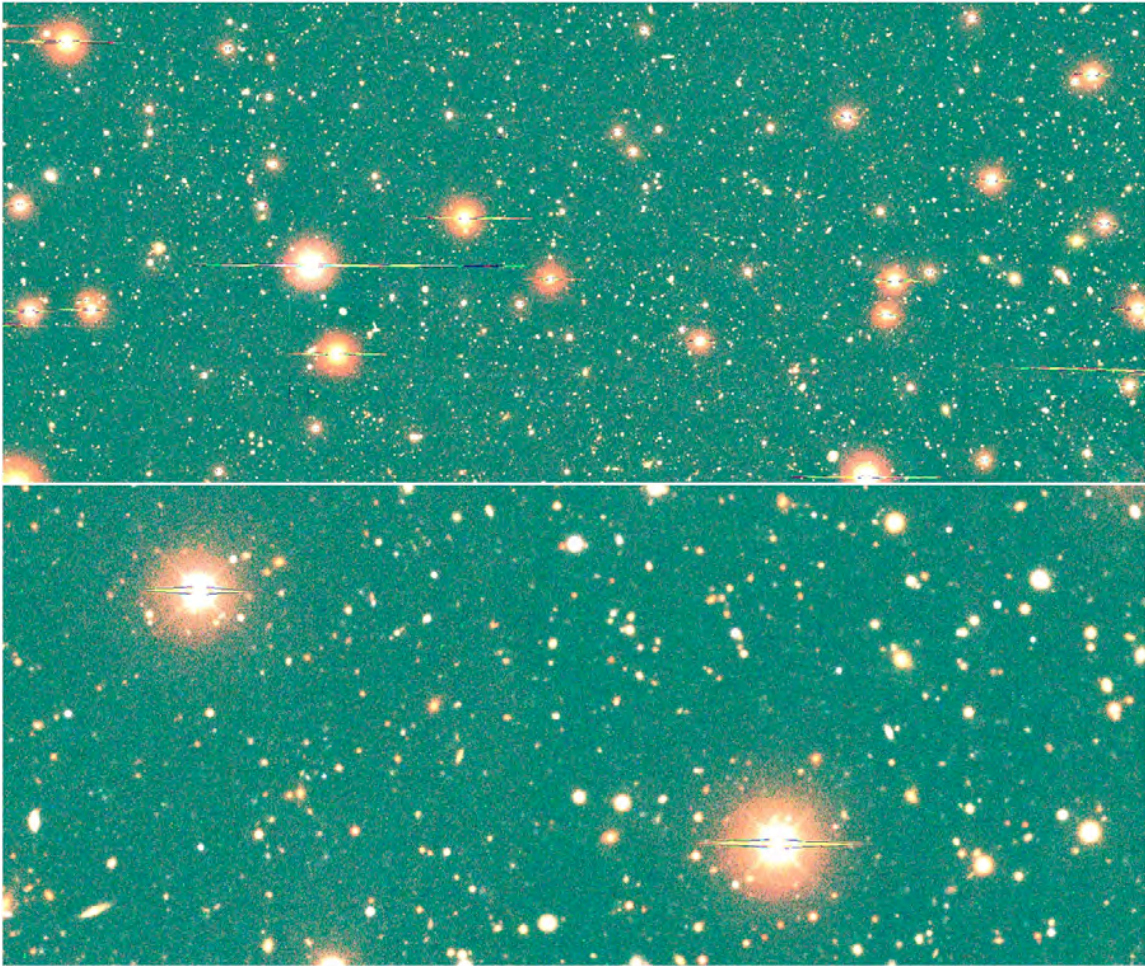

Figure 8: At top is an unmasked BCS[5] pseudo color image built using the *g*, *r*, and *i* band coadded images that span approximately 0.5°. At bottom is a zoom of a few arcminute region around the two central stars. The DESDM coadd employs SWarp for remapping and image combination, but the photometric zeropoints are determined using DESDM specific code. The scientific and aesthetic data quality is excellent.

### 3.2.2. Processing of BCS Data

Since DC3 we have used the DESDM system to process and calibrate the first three seasons of BCS[1] data. The BCS is an NOAO survey program that began in 2005 with the goal of providing early optical, multiband data to support the South Pole Telescope (SPT)[6] Sunyaev-Zel'dovich effect survey along with other mm wave imaging experiments like Atacama Pathfinder Experiment and the Atacama Cosmology Telescope. SPT serves as the initial and primary stimulus for DES, because without optical confirmation of SPT galaxy clusters together with cluster photometric redshifts and weak lensing mass constraints, the cosmological power of the SPT cluster cosmology experiment is dramatically reduced. With the goal of optical cluster confirmation and photometric redshift estimates for clusters out to z=1, BCS over 100 deg$^2$ can be thought of as a near-term mini-DES. It is a four band survey: *g*, *r*, *i* and *z*. The photometric depths in these bands are tuned so that we can extract galaxy photometric redshifts for L$_*$ and clusters galaxies to z=1. Because of the linearity bounds of the Mosaic2 detectors, achieving these depths requires a minimum of 10 Mosaic2 exposures in each portion of the survey: 2 exposures in *g* and *r* band and 3 exposures in *i* and *z*. With allocations made for the fourth and final BCS season in Fall 2008, CTIO has made 62 full nights on the Blanco 4m available. The full BCS list of collaborators is available on the collaboration website[5].

We processed the survey data, currently consisting of 50 nights of imaging on the Blanco 4m, using the Mercury cluster on TeraGrid. 5774 Mosaic2 exposures (0.75 TB) were processed into 24,696 reduced 2048x4096 images, 41,954 remapped images (transformed onto 36 arcminute by 36 arcminute tanget plane coadd tiles covering the survey) and 66,228 SExtractor catalogs of both reduced and remapped images. This corresponds to 1,165 deg$^2$ of science imaging data with 74.4 million cataloged and calibrated objects for a typical source density of around 9 objects/arcmin$^2$ in these single depth exposures.

In Figure 8 we show an example of the quality of the coadds with an unmasked coadded image spanning approximately 30 arcminutes across and a zoom into a central region of a few arcminutes. These pseudocolor images are produced using the coadd *g*, *r* and *i* band images in a particular tile within the survey. Figure 9 shows the galaxy counts in each band for a representative tile and the comparison of galaxy photometry across neighboring tiles whose photometry was calibrated independently. The quality of the astrometry and photometry is quite good to the target depths, although we are subject to the same biases in our Kron-like photometry for faint stars and galaxies as in the simulated DECam data. Work proceeds on completing all the coadd tiles and in estimating photometric redshifts over the full BCS survey. In summary, our testing shows that the DESDM system in its current state is already producing science ready data in extremely large simulated and quite large real optical imaging surveys.

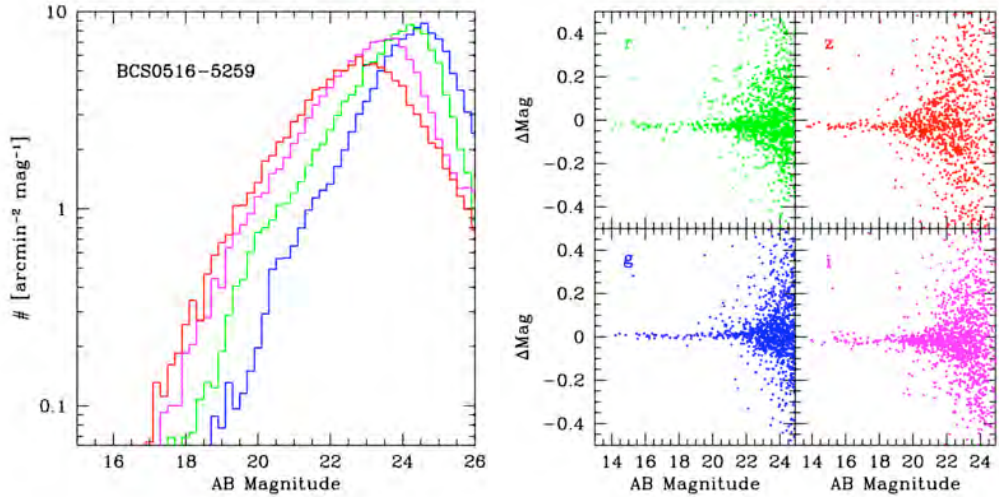

Figure 9: Galaxy sources counts and photometric error scatter plots for a representative coadd tile BCS0516-5259 in the BCS survey[5] that has been processed with the DESDM System. The turnovers in the galaxy counts occur a bit fainter than our targeted 10σ photometric depths of $g_{lim}$=24.0 (blue), $r_{lim}$=23.9 (green), $i_{lim}$=23.6 (magenta) and $z_{lim}$=22.3 (red). The photometric scatter plot shows comparisons of objects that appear in the 2 arcminute overlap with the independently calibrated neighboring coadd tile BCS0516-5332. The galaxy photometry agreement is excellent.

**4. Quality Assurance and Provenance**

The DESDM System design supports high levels of automation that require new approaches to assessing and assuring the quality of the data products. We are currently prototyping the DESDM QA framework with the goal of having a working version in place by the beginning of the next data challenge. The QA framework is designed to meet the data assessment needs, and it comprises (1) module level testing of data products, (2) module level processing status reports, (3) forwarding of QA and status information through an event channel to a central repository, (4) storage of data product QA as metadata within the DES Archive and (4) a monitoring portal that provides operators with summaries of the QA and status information produced at the module level as well as links to more detailed reports and the data themselves.

We have defined a range of status levels that allow modules to report a variety of runtime information ranging from the current data that the module is processing to module-failure messages associated with missing input data or filesystem access trouble. Failure-level status messages or data quality problems that rise above preset thresholds will trigger alerts that the DESDM operators can follow up through the QA portal. Detailed QA reports will be generated during nightly processing on each image and catalog, and these reports will be linked from the QA portal pages. In our last data challenge we used a quality checking module developed by the DESDM-Brazil team, and these tests went well. Image QA information on the mean seeing, PSF ellipticity and variation, and sky brightness are now routinely captured by the DESDM system and ingested into the DES Archive to enable science quality filtering on the images and cataloged objects.

Preliminary QA will also take place on the mountain, where each image will be examined using the same QA tests we employ at NCSA. On the mountain it is expected that the QA testing will only be carried out on a subset of the 62 CCD images from every exposure. The immediate QA feedback on the mountain will provide input to the Scheduling and Observing Tool that serves as a guide to the DES observers.

In addition to image data quality, our system will incorporate continuous testing of the processing pipelines. These tests will range from examining the contents of a catalog file to make sure an appropriate number of objects are detected to examining the accuracy of the star-galaxy separator by comparison to previous images taken of the same area. With the long term nature of the survey we can use established data norms that include the typical number of cosmic rays per image to trigger warnings when processing results differ significantly from the norm.

DES scientists must be able to track measurements of a particular object back through all the processing steps and to the original source DECam exposures. The design of the DESDM system supports this. Every object is linked to its image of origin using a unique ID, and this same approach is used to link from the coadd image of origin for the catalog through the intermediate images (including the input remapped images, the reduced images and the crosstalk corrected images) to the original exposure. System processes are all assigned unique run identifiers, and the binaries, configuration files, log files, xml pipelines, and svn branch of each relevant software package within the DESDM system are stored in the archive nodes along with the data products. The DES database is used to organize the linking of specific runs to characteristics of the software stack during that processing job, including the versions of the prerequisite libraries and other packages that the DESDM software stack is built within. Photometric calibration information for each object is stored so that the full history of photometric recalibration is available over the life of the survey and beyond. The provenance framework that we have constructed is sufficiently flexible that we can alter it to meet any additional requirements that emerge from the DES science working group tests of the DESDM processed data.

## 5. Integration of Science Codes

We have designed the DESDM system to enable the higher level science processing of the data. We are working with the science teams in weak lensing, cluster finding, photometric redshift estimation, supernovae standard candles and other areas to integrate scientific software into the DESDM system. This is important, because it will enable the broader collaboration and ultimately the public to more easily partake in the scientific analyses at the frontier without developing their own toolkits for, let's say, the lensing-analysis-ready morphological measurements or photometric redshift estimates. In cases where the higher level scientific processing benefits from a full DESDM-like operating environment, we are working directly with the relevant DES science working groups to co-develop and test pipelines. We provide an overview of the Weak Lensing Pipeline and the Difference Imaging Pipeline that we are beginning to develop.

### 5.1. Weak Lensing Pipeline

The weak lensing pipeline requires more integration than most of the other science software due to the nature of its data access. For the weak lensing science to succeed with the anticipated precision, we need to remove the effects of the point-spread function (PSF) anisotropy and scale to very high accuracy. One of the most difficult parts of this task involves characterizing the PSF at locations other than where there are observed stars. Stars give us a direct measurement of the PSF at the location of the star. But the PSF varies across the field, so interpolation to the locations of the galaxies is required. There are some indications that the interpolation can be made more accurate by utilizing the PSF information from many different exposures at once to determine appropriate interpolating functions for all of them.

A second difficulty inherent in the weak lensing analysis is that the shear estimates for each galaxy are best performed using all images of that galaxy at once in a single maximum likelihood solution. The solution involves a non-linear maximization, so there is no easy way to analyze each image separately and then combine the shear information after the fact. Therefore, the weak lensing analysis requires two passes through the data. First, the PSF is measured as a function of position in all of the exposures. Next, these data are used *in toto* to determine the interpolation functions for each exposure. We are exploring the use of a principle component analysis of the PSF and its dependencies on (1) instrument parameters such as CCD position and band, (2) telescope parameters such as hour angle, zenith angle and focus and (3) environmental parameters such as wind direction, temperature and wind speed. Then, for each galaxy, we input all the images of the galaxy along with the interpolated PSF for each of those exposures to find the single best estimate of the shear. In practice, we will do this latter step in chunks of area on the sky rather than on each single galaxy in a random-access fashion. Therefore, the image data access pattern has much in common with the data access for the coadd pipeline.

Dr. Jarvis is currently developing the joint image shear extraction code, and we can extract a model of the PSF directly using the PSFEx runs or by modeling that employs the shapelet expansion code that Dr. Sheldon delivered for testing within DC3. We plan to test the Weak Lensing Pipeline this fall on a simulated dataset that includes shear of the galaxy population.

### 5.2. Difference Imaging Pipeline

DESDM will work closely with Dr. Marriner and others in the SNe SWG to deploy a difference-imaging pipeline for processing the main DES survey dataset. The required algorithms are closely related to those we have developed for the PSF homogenization[7,8] within the coadd pipeline. The DES difference imaging pipeline will be developed using experience gained in the development of the difference imaging pipeline currently in use within the SDSS supernovae project. We expect that the difference imaging pipeline used for the main survey to detect and catalog variable objects will be very similar to the primary difference imaging pipeline used within the SNe specific DES fields that are reobserved with a rapid cadence. Joint development work has just begun, and we are hoping to test an early prototype of the DES pipeline in DC4 this fall.

### 6. Conclusion

The DESDM system is a modern astronomical data management system for the processing of DES data that leverages publicly available high performance computing (HPC) resources on TeraGrid and the Open Science Grid. An HPC data management system offers the flexibility to support the daily survey processing while at the same time supporting the two orders of magnitude larger seasonal reprocessing. Our approach offers significant hardware and hardware maintenance savings, because the project must only purchase DES specific hardware such as database servers, web servers and data storage on disk and on tape.

The DESDM system exists in prototype form and has been successfully used in the automated processing and calibration of 8000 deg$^2$ of simulated DECam data and over 1100 deg$^2$ of Mosaic2 data from the Blanco Cosmology Survey (BCS). The photometric and astrometric quality of the resulting data is excellent, and higher level science analyses of the BCS data are proceeding. Additional improvements are underway to meet the 2% photometry requirement of the DES.

The DESDM development project is proceeding through yearly cycles of development and testing using extensive datasets of increasing scale and realism. A final stress test on a full season of simulated data is scheduled in Fall 2010, and the DESDM system along with DECam will be commissioned in early 2011. We expect the first season of science observations to being in Fall 2011, and the survey will continue for 525 nights extending over five observing seasons. The source and detrended DES image data will be released to the public one year after they are acquired, and the higher level science data products will be released once in the middle of the survey and again one year after the completion of the survey. DES collaboration science analyses will proceed through a number of science working groups. Our goal is to deliver four powerful and independent constraints on the nature of the cosmic acceleration using (1) a galaxy cluster survey, (2) weak lensing, (3) galaxy clustering and (4) Type Ia supernova distance measurements.

The DESDM system includes features that would very likely benefit other data and compute intensive experiments, and so we are working toward making it easy to use and available. We will release the DESDM system to the community by providing access to our software repository. In addition, we will make available a DECam reduction portal that will enable non-DES users of DECam to process their data within the standard DESDM processing environment.

## ACKNOWLEDGEMENT

The DESDM team acknowledges continuing support from NSF AST 07-15036 as well as significant seed funding provided by the National Center for Supercomputing Applications and the University of Illinois Department of Astronomy, the College of Language Arts and Science, and the Vice Chancellor for Research.

Funding for the DES Projects has been provided by the U.S. Department of Energy, the U.S. National Science Foundation, the Ministry of Science and Education of Spain, the Science and Technology Facilities Council of the United Kingdom, the National Center for Supercomputing Applications at the University of Illinois at Urbana-Champaign, the Kavli Institute of Cosmological Physics at the University of Chicago, Financiadora de Estudos e Projetos, Fundação Carlos Chagas Filho de Amparo à Pesquisa do Estado do Rio de Janeiro , Conselho Nacional de Desenvolvimento Científico e Tecnológico and the Ministério da Ciência e Tecnologia and the Collaborating Institutions in the Dark Energy Survey.

The Collaborating Institutions are Argonne National Laboratories, the University of Cambridge, Centro de Investigaciones Energeticas, Medioambientales y Tecnologicas-Madrid, the University of Chicago, University College London, DES-Brazil, Fermilab, the University of Edinburgh, the University of Illinois at Urbana-Champaign, the Institut de Ciencies de l'Espai (IEEC/CSIC), the Institut de Fisica d'Altes Energies, the Lawrence Berkeley National Laboratory, the University of Michigan, the National Optical Astronomy Observatory, the Ohio State University, the University of Pennsylvania, the University of Portsmouth and the University of Sussex.